\documentclass[prb,aps,twocolumn,showpacs]{revtex4}
\usepackage{graphicx,latexsym}
\usepackage{dcolumn}
\usepackage{amsmath,amssymb,epsf,bm}
\begin{document}
\title{Spin-Hall interface resistance in terms of Landauer type spin dipoles}
\author{A.~G. Mal'shukov$^1$, L.~Y. Wang$^{2}$, C.~S. Chu$^{2}$}
\affiliation{$^1$Institute of Spectroscopy, Russian Academy of
Science, 142190, Troitsk, Moscow oblast, Russia \\
$^2$Department of Electrophysics, National Chiao Tung University,
Hsinchu 30010, Taiwan}
\begin{abstract}
We considered the nonequlibrium spin dipoles induced around spin
independent elastic scatterers by the intrinsic spin-Hall effect
associated with the Rashba spin-orbit coupling. The normal to 2DEG
spin polarization has been calculated in the diffusion range
around the scatterer. We found that although around each impurity
this polarization is finite, the corresponding macroscopic spin
density, obtained via averaging of individual spin dipole
distributions over impurity positions is zero in the bulk. At the
same time, the spin density is finite near the boundary of 2DEG,
except for a special case of a hard wall boundary, when it turns
to 0. The boundary value of the spin polarization can be
associated with the interface spin-Hall resistance determining the
additional energy dissipation due to spin accumulation.
\end{abstract}
\pacs{72.25.Dc, 71.70.Ej, 73.40.Lq}

\maketitle
\section{Introduction}
Most of the theoretical studies on the spin-Hall effect (SHE) has
been devoted to calculation of the spin current (for a review see
Ref. \onlinecite{Engel}). Such a current is a linear response to
the external electric field $\mathbf{E}$ which induces a spin flux
of electrons or holes flowing in the direction perpendicular to
$\mathbf{E}$. This spin flux can be due either to the intrinsic
spin-orbit interaction (SOI) inherent to a crystalline solid
\cite{Murakami}, or to spin dependent scattering from impurities
\cite{Hirsch}. The spin-Hall current, as a response to the
electric field, is characterized by the spin-Hall conductivity. On
the other hand, similar to the conventional Hall effect, one can
introduce the spin-Hall resistivity from calculating the local
chemical potential difference $\mu_s=\mu\uparrow-\mu\downarrow$ in
a response to the DC electric current. For 2DEG in a local
equilibrium this potential difference can be related to the
z-component (perpendicular to 2DEG) of the spin polarization,
according to $S_z=N_F\mu_s$, where $N_F$ is the density of states
near the Fermi level. Therefore, the spin-Hall resistivity is
closely associated with spin accumulation near interfaces. It
should be noted that measuring spin polarization is thus far the
only realistic way to detect SHE \cite{Wunderlich,Awschalom2}. For
interfaces of various nature such an accumulation has been
calculated in a number of works
\cite{MalshStrip,Accumulation2,Accumulation4,Accumulation5,Accumulation1,Accumulation3,ballistic}.
A typical example to study spin accumulation is an infinite along
$x$-direction 2D strip with a width $w$ along $y$-direction. In
this geometry the DC current flows in x-direction, while the
spin-Hall current flows in y-direction with the spin density
accumulating near boundaries. An analog of the Hall voltage could
be a difference of $\mu_s$ on both sides of the strip. There is,
however, a fundamental distinction from the charge Hall effect. In
the latter case, due to the long-range nature of the electric
potential created by conserving electric charges, the Hall voltage
is proportional to the width of the strip. In contrast, the
spin-Hall electrochemical potential at the interface does not
depend on $w$ as $w\rightarrow\infty$ because spin relaxation
essentially suppresses the long range contribution to spin
polarization buildup near interfaces. Hence, it is sensible to
introduce an interface spin-Hall resistance, which is the
proportionality coefficient between the interface value of $\mu_s$
and the electric current density.

Below, we will consider the spin-Hall resistance from the
microscopic point of view. This approach is based on the
Landauer's \cite{Landauer} idea that at a given electric current
each impurity is surrounded by a nonequlibrium charge cloud
forming a dipole. Combined together these dipoles create a voltage
drop across the sample. Therefore, each impurity plays a role of
an elementary resistor. In a similar way, nonequlibrium spin
dipoles could be induced subsequent to the spin-Hall current. One
may expect that the spin cloud will appear around a spin-orbit
scatterer in case of extrinsic SHE, as well as around a
spin-independent scatterer, in case of the intrinsic effect. The
latter possibility for a 2D electron gas with Rashba interaction
has been considered in Ref. \onlinecite{MalshCloud}. The
perpendicular to 2DEG polarization was calculated within the
ballistic range around a scatterer. On the other hand, in order to
study spin accumulation and the spin-Hall resistance on a
macroscopic scale, one needs to calculate the spin density
distribution at distances much larger than the mean free path $l$
of electrons. Below, we will extend the Green function method of
Ref. \onlinecite{MalshCloud} to the diffusive range. In Section II
the spin density distribution around an individual target impurity
will be calculated. In Section III we will consider the interface
spin accumulation created by spin dipoles randomly but
homogeneously distributed in space. A relation between spin-Hall
resistance and energy dissipation will be discussed in Sec. IV. A
summary and discussion of results will be presented in Section V.

\section{Spin cloud induced by a single impurity}

As known, the electric field applied to a homogeneous 2DEG with
Rashba SOI induces a parallel to 2DEG component of the
nonequlibrium spin polarization \cite{Edelstein}. The spin-Hall
effect produces, however, a zero spin polarization in its
z-component. This understanding about such a \emph{homogeneous}
gas has implied an averaging over impurity positions. An impure
system, on the other hand, can not be uniform on a microscopic
scale. The effect of each impurity on the spin polarization could
be singled out by considering an impurity (a target impurity) at a
fixed position while taking at the same time the average over
positions of other impurities. In such a way the Landauer electric
dipole has been calculated \cite{Chu,Zwerger}. The electron
density around a target impurity represented by the elastic
scatterer was found from the asymptotic expansion of the scattered
wave functions of electrons. At the same time, wavevectors of
incident particles were weighted with the nonequlibrium part of
the Boltzmann distribution function. We will employ another method
based on the Green function formalism \cite{MalshCloud}. Within
this method the spin density response to the electric field
$\mathbf{E}$ is given by the standard Kubo formula with the
scattering potential of the target impurity incorporated into the
retarded and advanced Green functions
$G^{r/a}(\mathbf{r},\mathbf{r}^{\prime},\omega)$ denoted by the
superscripts $r$ and $a$, respectively. As such, the n-component
of the stationary spin polarization is given by
\begin{eqnarray}\label{sigma}
S_n(\mathbf{r})&=-\frac{e}{m^{*}}&\int d^2 r^{\prime}\int
\frac{d\omega}{2\pi} \frac{d n_F(\omega)}{d\omega} \times
\nonumber
\\
&& Tr[\overline{\sigma^n
G^r(\mathbf{r},\mathbf{r}^{\prime},\omega)(\mathbf{v}\mathbf{E})G^a(\mathbf{r}^{\prime},\mathbf{r},\omega)}]
\,,
\end{eqnarray}
where the overline denotes averaging over impurity positions, the
trace runs through the spin variables and $n_F(\omega)$ is the
Fermi distribution function. To avoid further confusion we note
that the angular moment is obtained by multiplying
$S_n(\mathbf{r})$ by $\hbar/2$ and $e$ is the particle charge,
which is negative for electrons. At low temperatures only $\omega$
in close vicinity around $E_F$ contributes to the integral in
(\ref{sigma}). Therefore, below we set $\omega=E_F$ and omit the
frequency argument in the Green functions. Further, $\mathbf{v}$
is the particle velocity operator containing a spin dependent part
associated with SOI. Writing SOI in the form
\begin{equation}\label{Hso}
H_{so} = \mathbf{h}_{\mathbf{k}}\cdot\bm{\sigma} \, ,
\end{equation}
one obtains the velocity operator
\begin{equation}\label{v}
\mathrm{v}^j=\frac{k^j}{m^*}+\frac{\partial
\mathbf{h}_{\mathbf{k}}\cdot \bm{\sigma}}{\partial k^j}\,,
\end{equation}
where $\bm{\sigma}$$\equiv$$(\sigma^x,\sigma^y,\sigma^z)$ is the
Pauli matrix vector. In case of Rashba interaction the spin-orbit
field $\mathbf{h}_{\mathbf{k}}$ is given by
\begin{equation}\label{hRashba}
h_x=\alpha k_y \,\,\,,\,\,\, h_y=-\alpha k_x \,.
\end{equation}
We assume that the target impurity, located at $\mathbf{r}_i$, is
represented by a scattering potential
$U(\mathbf{r}-\mathbf{r}_i)$. The Green functions in (\ref{sigma})
have to be expanded in terms of this potential. Up to the second
order in $U$ one obtains
\begin{widetext}
\begin{eqnarray}\label{G}
G^{r/a}(\mathbf{r},\mathbf{r'})&=&G^{r/a(0)}(\mathbf{r},\mathbf{r'})+\int
ds^2
G^{r/a(0)}(\mathbf{r},\mathbf{s})U(\mathbf{s}-\mathbf{r}_i)G^{r/a(0)}(\mathbf{s},\mathbf{r'})+
\nonumber \\ && \int ds^2ds^{\prime \, 2}
G^{r/a(0)}(\mathbf{r},\mathbf{s})U(\mathbf{s}-\mathbf{r}_i)G^{r/a(0)}(\mathbf{s},\mathbf{s'})
U(\mathbf{s'}-\mathbf{r}_i)G^{r/a(0)}(\mathbf{s'},\mathbf{r'})\,.
\end{eqnarray}
\end{widetext}
The unperturbed functions $G^{r/a(0)}$ depend, nevertheless, on
scattering from background random impurities. The latter create
the random potential $V_{sc}(\mathbf{r})$ which is assumed to be
delta correlated, so that the pair correlator $\langle
V_{sc}(\mathbf{r})V_{sc}(\mathbf{r}^{\prime})\rangle=\Gamma
\delta(\mathbf{r}-\mathbf{r}^{\prime})/\pi N_F$, where
$\Gamma=1/2\tau$ is expressed via the mean elastic scattering time
$\tau$. The delta correlation means that the corresponding
impurity potential is the short range one. In fact, the potential
of the target impurity could be different from that of the random
impurities. It might be a special sort of impurities added to the
system. On the other hand, the target and the random impurities
would be identical if one would try to employ the spin dipoles for
the interpretation of spin accumulation near interfaces.
\begin{figure}[tp]
\includegraphics[width=7.5cm, height=6cm]{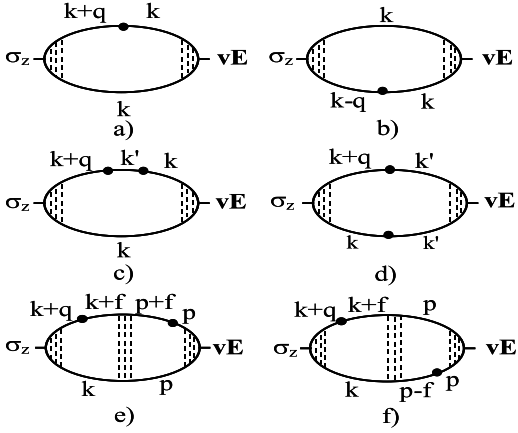}
\caption {Examples of diagrams for the spin density. Scattering of
electrons by a target impurity is shown by the solid circles. Dashed
lines denote the ladder series of particle scattering by the random
potential. $\mathbf{p}, \mathbf{k}, \mathbf{k'}$ are electron
momenta.} \label{fig1}
\end{figure}

After substitution of (\ref{G}) into (\ref{sigma}) one must
calculate background impurity configurational averages containing
products of several Green functions $G^{(0)}$. Assuming that the
semiclassical approximation $E_F\tau \gg 1$ is valid, the standard
perturbation theory \cite{agd,alt} can be employed whose building
blocks are the so called ladder perturbation series expressed in
terms of the unperturbed average Green functions
\begin{equation}\label{Gk}
 G^{r/a}_{\mathbf{k}}=\int
 d^2(\mathbf{r}-\mathbf{r}^\prime)e^{i\mathbf{k}(\mathbf{r}-\mathbf{r}^\prime)}
 \overline{G^{r/a(0)}(\mathbf{r},\mathbf{r'})}
\end{equation}
given by the 2$\times$2 matrix
\begin{equation}\label{G0}
 G^{r/a}_{\mathbf{k}}
 = (E_F - E_{\mathbf{k}} -
\bm{h}_{\mathbf{k}}\cdot\bm{\sigma} \pm i\Gamma)^{-1} \, ,
\end{equation}
where $E_{\mathbf{k}}$=$k^2/(2m^*)$. When averaging the Green
function products, within the ladder approximation only pairs of
retarded and advanced functions carrying close enough momenta
should be chosen to become elements of the ladder series. After
decoupling the mean products of Green functions into the ladder
series, the Fourier expansion of (\ref{sigma}) can be represented
by diagrams shown at Fig.1. In these diagrams the diffusion ladder
renormalizes both the lefthand and righthand vertices. The
renormalized lefthand vertex $\Sigma_z(\mathbf{q})$ is associated
with the $\mathbf{q}$-th Fourier component of the induced spin
density, and the corresponding diffusion propagator enters with
the wavevector $\mathbf{q}$. In its turn, the righthand vertex
$T(\mathbf{p})$ related to the homogeneous electric field, is
represented by the ladder at the zeroth wavector. The
corresponding physical process is the D'yakonov-Perel spin
relaxation of a uniform spin distribution. This lefthand vertex
alone contributed to the ballistic case result \cite{MalshCloud},
while $\Sigma_z(\mathbf{q})$ has been taken unrenormalized  due to
large values of $\mathbf{q} \gg 1/(v_F \tau)1$ in the ballistic
regime. Figs.1(e),1(f) represent some diagrams where the diffusion
process separates two scattering events. As it will be shown
below, such diagrams give rise to small corrections to the spin
density, and can be neglected. Hence, the main contribution comes
from diagrams similar to those in Figs.1(a)-1(d). The
corresponding spin polarization has the form
\begin{equation}\label{sigma2}
S_z(\mathbf{q})=\frac{1}{2\pi}\sum_{\mathbf{p},\mathbf{k}}
Tr[G^a_{\mathbf{p}\mathbf{k}} \Sigma_z(\mathbf{q})
G^r_{\mathbf{k}+\mathbf{q},\mathbf{p}} T(\mathbf{p})] \,.
\end{equation}
The functions $G^{r/a}_{\mathbf{k}^{\prime}\mathbf{k}}$ are
formally represented by the Fourier expansion of Eq.~(\ref{G})
with respect to $\mathbf{r}$ and $\mathbf{r}^{\prime}$, providing
that the respective average values
$\overline{G^{(0)}(\mathbf{r},\mathbf{r'})}$ are substituted
instead of $G^{(0)}(\mathbf{r},\mathbf{r'})$. Evaluating the pair
products of such functions in (\ref{sigma2}), one should take into
account only terms up to the second order with respect to the
scattering potential $U$.

The vertices $\Sigma_z(\mathbf{q})$ and $T(\mathbf{p})$ can be
easily calculated. As it was discussed in Ref.
\onlinecite{MalshCloud}, due to considerable cancellation of
diagrams which is known from literature on the spin-Hall effect,
$T(\mathbf{p})$ acquires a quite simple form in a special case of
Rashba SOI. Namely,
\begin{equation}\label{TRashba}
T(\mathbf{p})=\frac{e}{m^*}\mathbf{p}\cdot\mathbf{E} \,.
\end{equation}
In its turn, $\Sigma_z(\mathbf{q})$ is expressed in terms of the
diffusion propagator. Indeed, let us represent this vertex using a
basis of four 2$\times$2 matrices $\tau^0=1$ and $\tau^i=\sigma^i$
with $i=x,y,z$. Then, $\Sigma_z(\mathbf{q})$ can be written as
\begin{equation}\label{Sigma}
\Sigma_z(\mathbf{q})=\sum_b D^{zb}(\mathbf{q}) \tau^b\,\,,\,\,\,
b=0,x,y,z
\end{equation}
where $D^{zb}(\mathbf{q})$ are the matrix elements of the
diffusion propagator satisfying the spin diffusion equation, as it
was described in Ref. \onlinecite{MalshStrip} and references
therein. The nondiagonal element $D^{z0}(\mathbf{q})$ appears due
to the spin-charge mixing and it is zero for SOI of quite general
form, including the Rashba interaction
\cite{MalshDress,Mischenko,Burkov}. Finally, from (\ref{sigma2}),
using (\ref{TRashba}) and (\ref{Sigma}), we express
$S_z(\mathbf{q})$ in the form
\begin{equation}\label{sigma3}
S_z(\mathbf{q})=\sum_{n=x,y,z}
D^{zn}(\mathbf{q})I^n(\mathbf{q})\,,
\end{equation}
where $I^n(\mathbf{q})$
\begin{equation}\label{I}
I^n(\mathbf{q})=\frac{e}{2\pi
m^*}\sum_{\mathbf{p},\mathbf{k}}(\mathbf{p}\cdot\mathbf{E})Tr[G^a_{\mathbf{p}\mathbf{k}}
\sigma^n G^r_{\mathbf{k}+\mathbf{q},\mathbf{p}} ]\,.
\end{equation}
The function $I^n(\mathbf{q})$ has a simple physical meaning. For
$n=x,y,z$ it represents a source of spin polarized particles
emitting from the target impurity. Their further diffusion and
spin relaxation result in the observable polarization. This source
term feature is conceptually similar, though different in its
context, to the original charge cloud consideration when SOI is
not present and Boltzmann equation is used to describe the
subsequent background scattering \cite{Landauer,Chu}. For $q \ll
l^{-1} \ll k_F$ the source can be expanded in powers of $q$.
Therefore, the wavevector independent terms represent the delta
source located at $\mathbf{r}_i$, while the linear in $q$ terms
are associated with the gradient of the delta-function. Below we
will keep only the constant and linear terms for each n-th
component $I^n(\mathbf{q})$ and assume, for simplicity, the short
range scattering potential $U(\mathbf{r})$, so that its
$\mathbf{k}$-th Fourier transform is simply
$U\exp(-i\mathbf{k}\cdot\mathbf{r}_i)$, where $U$ is a constant.
Further, $I^n(\mathbf{q})$ can be written as
\begin{equation}\label{I12}
I^n(\mathbf{q})=I^n_1(\mathbf{q})+I^n_2(\mathbf{q}) \,,
\end{equation}
where $I_1$ and $I_2$ are of the first and the second order with
respect to the scattering potential $U$, respectively.
Accordingly, $I_1$ and $I_2$ are represented by Figs. 1(a),1(b)
and Figs. 1(c),1(d), respectively. Using (\ref{G}) to express
Green functions $G^{r/a}_{\mathbf{k}^{\prime}\mathbf{k}}$ in
(\ref{I}) we obtain
\begin{eqnarray}\label{I1}
I^n_1(\mathbf{q})&=&\frac{eU}{2\pi
m^*}e^{i\mathbf{q}\cdot\mathbf{r}_i}\sum_{\mathbf{p}}(\mathbf{p}\cdot\mathbf{E})
 \times
\nonumber \\
&&Tr[G^r_{\mathbf{p}}G^a_{\mathbf{p}}\left(\sigma^n
G^r_{\mathbf{p+q}}+ G^a_{\mathbf{p-q}}\sigma^n\right)] \,,
\end{eqnarray}
and
\begin{eqnarray}\label{I2}
&&I^n_2(\mathbf{q})=\frac{eU^2}{2\pi
m^*}e^{i\mathbf{q}\cdot\mathbf{r}_i}\sum_{\mathbf{p}\mathbf{k}}(\mathbf{p}\cdot\mathbf{E})
Tr[G^r_{\mathbf{p}}G^a_{\mathbf{p}}\times
\nonumber \\
&&\left(G^a_{\mathbf{k}}\sigma^n G^r_{\mathbf{k+q}}
-\gamma\sigma^n
 G^r_{\mathbf{p+q}}+ \gamma
G^a_{\mathbf{p-q}}\sigma^n\right)] \,,
\end{eqnarray}
where
\begin{equation}\label{gamma}
\gamma=i\mathrm{Im}[\sum_{\mathbf{k}}G^a_{\mathbf{k}}]=i\pi N_F
\end{equation}
In our following consideration we let the x-axis to be parallel
with the electric field, and the z-axis to be perpendicular to the
2DEG. The system Hamiltonian is symmetric under a symmetry
operation combining a reflection from the plane perpendicular to
the y-axis, that means $p_y \rightarrow -p_y$, and a unitary
transformation $\sigma^i \rightarrow \sigma_y\sigma^i\sigma_y$.
Applying this transformation to (\ref{I}) one can easily see that
$I^x(q_x,q_y)=-I^x(q_x,-q_y)$, $I^z(q_x,q_y)=-I^z(q_x,-q_y)$ and
$I^y(q_x,q_y)=I^y(q_x,-q_y)$. Making use of another symmetry
operation $p_x \rightarrow -p_x$, $p_y \rightarrow -p_y$ and
$\sigma^i \rightarrow \sigma_z\sigma^i\sigma_z$, we obtain
$I^x(q_x,q_y)=I^x(-q_x,-q_y)$, $I^z(q_x,q_y)=-I^z(-q_x,-q_y)$ and
$I^y(q_x,q_y)=I^y(-q_x,-q_y)$. From these relations it is easy to
see that expansion of $I^z$ into power series starts from linear
in $\mathbf{q}$ terms, while the leading term in $I^y$ is $const$
and the next one is quadratic in $q$. On this reason only $const$
will be taken into account in $I^y$. The expansion of $I^x$ starts
from $q_xq_y$, and this source component will be neglected.

Calculation of $I_1$ and $I_2$ given by Eqs. (\ref{I1}),(\ref{I2})
is based on the standard linearization near the Fermi level, thus
ignoring band effects giving rise to small corrections $\sim
h_{k_F}/E_F, \Gamma/E_F$. Further, the diffusion approximation is
valid at $q \ll 1/l$. At the same time, the characteristic
lengthscale is determined by the spin relaxation length $l_{so}$,
which is the distance a particle diffuses during the
D'yakonov-Perel' spin relaxation time
$\tau_{so}=4(h_{k_F}^2\tau)^{-1}$. The corresponding diffusion
length $l_{so}=\sqrt{D\tau_{so}}$, where $D=v_{F}^2 \tau/2$ is the
diffusion constant. Hence, $l_{so}=v_F/h_{k_F}$. Taking $q \sim 1/
l_{so}$ one finds that the diffusion approximation is valid if
$h_{k_F}/\Gamma \ll 1$. Therefore, within this approximation we
will retain only the leading powers of $h_{k_F}/\Gamma \ll 1$. In
such a way, direct calculation of $I^n_1$ with the Green functions
and SOI given by Eqs. (\ref{G0}) and (\ref{hRashba}),
respectively, shows that both $I^y_1$ and $I^z_1$ are small by a
factor $\Gamma/E_F$. For example, using the relation
\begin{equation}\label{derivative}
(G^{r/a}_{\mathbf{k}})^2=-\frac{\partial}{\partial
E_F}G^{r/a}_{\mathbf{k}}
\end{equation}
which follows from (\ref{G0}), evaluating $I^y_1$ at $q=0$, one
can represent the corresponding sum in (\ref{I1}) as
\begin{eqnarray}\label{derivative2}
-\frac{\partial}{\partial E_F}\sum_{\mathbf{p}} p_x
Tr[G^r_{\mathbf{p}}G^a_{\mathbf{p}}\sigma^y]=\nonumber\\
-\frac{\partial}{\partial E_F}\left( \frac{2\pi}{\Gamma} N_F
m^{*}\frac{\partial h^y_{\mathbf{p}}}{\partial p_x}\right)\,.
\end{eqnarray}
In case of Rashba SOI with the constant coupling strength $\alpha$
and energy independent parameters $\Gamma, m^{*}$ and $N_F$, the
sum (\ref{derivative2}) is equal to 0. Otherwise, it is finite,
but small due to the smooth energy dependence of these parameters.
Similar analysis, although not so straightforward, can be applied
to $I^z_1$, which is linear in $q$. The smallness of $I^z_1$ can
be also seen from Ref. \onlinecite{MalshCloud} where the linear in
$U$ contribution to the spin density was associated with fast
Friedel oscillations. It is clear that their Fourier transform
will be small in the range of $q \ll k_F^{-1}$.

At the same time  $I^y_2$ and $I^z_2$ are not zero. They are given
by
\begin{eqnarray}\label{I2fin}
I^y&=&v_d N_F m^*\alpha h_{k_F}^2\frac{\Gamma^{\prime}}{\Gamma^{3}}  \nonumber\\
I^z&=&-iq_y v_d N_F h_{k_F}^2\frac{\Gamma^{\prime
}}{2\Gamma^{3}}\nonumber\\
I^x&=&0 \,,
\end{eqnarray}
where $\Gamma^{\prime}=\pi N_F U^2$ and $v_d=eE\tau/m^*$ is the
electron drift velocity. If the target impurity is represented by
one of the random scatterers, we get $\Gamma^{\prime}=\Gamma/n_i$,
where $n_i$ is the density of impurities.

In the above calculation we did not take into account the diagrams
shown in Figs. 1(e)-1(f) and those similar to them. It can be
easily seen that such diagrams contain $I^n_1$ as a factor. For
example, the sum of diagrams at Fig. 1(e)-1(f) contains as a
multiplier the sum of diagrams shown at Figs. 1(a)-1(b).
Therefore, such diagrams are small by the same reason as $I^n_1$
are, at least, in the most important range of $f \ll l^{-1}$.
Particularly in this range of small $f$ the diffusion propagator
between the two scattering events in Figs. 1(e)-1(f) becomes
large.

Now one can combine the source $I^n$ with the diffusion propagator
to find from Eq. (\ref{sigma3}) the shape of the spin cloud around
a single scatterer. Taking into account (\ref{I2fin}), Eq.
(\ref{sigma3}) is transformed into
\begin{equation}\label{sigma4}
S_z(\mathbf{q})= -v_d N_F h_{k_F}^2\frac{\Gamma^{\prime
}}{2\Gamma^{3}}\left[iq_y D^{zz}(\mathbf{q})-2m^*\alpha
 D^{zy}(\mathbf{q})\right] \,.
\end{equation}
The matrix elements $D^{ij}(\mathbf{q})$ satisfy the spin
diffusion equation \cite{MalshStrip,MalshDiff}
\begin{equation}\label{diffusion}
\sum_{l}\left(-\delta^{il} D\bm q^2-\Gamma^{il}+i\sum_m R^{ilm}q_m
\right)D^{lj}(\mathbf{q})=-2\Gamma\delta_{ij} \,,
\end{equation}
where the matrix $\Gamma^{il}$ determining the D'yakonov-Perel'
spin relaxation rates is given by
\begin{equation}\label{DP}
\Gamma^{il} = 4\tau \, \langle\delta^{il} h^2_{\bm{k}_F} -
h^i_{\bm{k}_F} h^l_{\bm{k}_F}\rangle \,,
\end{equation}
with angular brackets denoting averaging over the Fermi surface.
In the case of Rashba SOI, Eq. (\ref{hRashba}), one gets
$\Gamma^{zz}=4\tau h_{k_F}^2$ and $\Gamma^{xx}=\Gamma^{yy}=2\tau
h_{k_F}^2$. The last term in lhs of (\ref{diffusion}) is
associated with spin precession in SOI field. It has the form
\begin{equation}\label{precession}
R^{ilm}=4\tau\sum_p\varepsilon^{ilp}\, \langle h_{\bm{k}}^{p}
v_F^m\rangle \,.
\end{equation}
For the Rashba SOI the nonzero components are
\begin{equation}\label{R}
i\sum_m R^{izm}q_m=-i\sum_m R^{zim}q_m= 4iDm^*\alpha q_i \,.
\end{equation}
We ignored in (\ref{diffusion}) a small term which gives rise to
the spin-charge mixing \cite{MalshStrip,Mischenko,Burkov}. This
mixing is already taken into account in the source term because
$I^n$ for $n=x,y,z$ describes the source of the \emph{spin
polarization} in response to the \emph{electric} field. From Eqs.
(\ref{diffusion})-(\ref{R}) one finds
\begin{eqnarray}\label{D}
D^{zz}&=&\frac{1}{2h^2_{\bm{k}_F}\tau^2}\frac{\tilde{q}^2+1}{(\tilde{q}^2+2)(\tilde{q}^2+1)-4\tilde{q}^2}  \nonumber\\
-D^{zy}&=&D^{yz}=\frac{1}{2h^2_{\bm{k}_F}\tau^2}\frac{2i\tilde{q}_y}{(\tilde{q}^2+2)(\tilde{q}^2+1)-4\tilde{q}^2}\nonumber\\
D^{yy}&=&\frac{1}{2h^2_{\bm{k}_F}\tau^2}\frac{\tilde{q}^2+2}{(\tilde{q}^2+2)(\tilde{q}^2+1)-4\tilde{q}^2}\,,
\end{eqnarray}
where $2\tilde{q}=l_{so}q$ denotes the dimensionless wavector.
Substituting (\ref{D}) into (\ref{sigma4}) we finally find
\begin{equation}\label{sigmafin}
S_z=-2iv_d \frac{m^* \alpha}{\hbar} N_F \frac{\Gamma^{\prime
}}{\Gamma}\frac{\tilde{q}_y(\tilde{q}^2+3)}{(\tilde{q}^2+2)(\tilde{q}^2+1)-4\tilde{q}^2}\,,
\end{equation}
and
\begin{equation}\label{sigmay}
S_y=2v_d \frac{m^* \alpha}{\hbar} N_F \frac{\Gamma^{\prime
}}{\Gamma}\frac{(3\tilde{q}^2+2)}{(\tilde{q}^2+2)(\tilde{q}^2+1)-4\tilde{q}^2}\,.
\end{equation}
To restore the conventional units we added $\hbar$ into
(\ref{sigmafin})-(\ref{sigmay}). The $z$ component of the spin
density in real space is shown in Fig.2. According to expectations
it has the shape of a dipole oriented perpendicular to the
electric field. Its spatial behavior is determined by the single
parameter $l_{so}$, which gives the range of exponential decay of
the spin polarization with increasing distance from an impurity.
The $S_y$ component averaged over impurity positions gives the
uniform bulk polarization. It is interesting to note that when the
target impurities are identical to background ones
($\Gamma^{\prime}=\Gamma$), the so obtained uniform polarization
$S_y|_{q\rightarrow 0}$ coincides with the electric spin
orientation \cite{Edelstein} $S_y=2v_d m^* \alpha N_F/\hbar$.
\begin{figure}[tp]
\includegraphics[width=7.5cm, height=6cm]{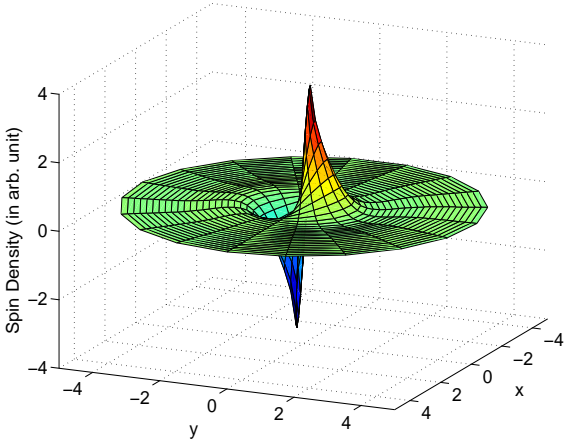}
\caption {Spatial distribution of $S_z$ component of the spin
density around a single scatterer. The unit of length $=l_{so}$}
\label{fig2}
\end{figure}

\section{Spin accumulation in a semiinfinite system}

In this section we will consider a semiinfinite electron gas $y>0$
bounded at $y=0$ by a boundary parallel to the electric field. Our
goal is to calculate a combined effect of spin clouds from random
impurities. It is important to note that summation of spin dipoles
from many scatterers does not result in a magnetic potential
gradient in the bulk of the sample. This is principally different
from the Landauer charge dipoles which are associated with the
macroscopic electric field. The origin of such a distinction can
be immediately seen from (\ref{sigmafin}). The magnetic potential,
as it was defined in Sec.I, is proportional to $S_z$. Taking its
gradient one gets $q_y S_z$. After averaging over impurity
positions $q \rightarrow 0$, so that $q_y S_z \rightarrow 0$. It
happens due to spin relaxation, which provides at $q =0$ a finite
value of the denominator in (\ref{sigmafin}). At the same time, in
case of the charge cloud, the denominator of the particle
diffusion propagator is proportional to $q^2$. Hence, the
corresponding gradient of the electrochemical potential (electric
field) is finite at $q=0$. Although the bulk magnetic potential is
zero, one can not expect that it will also be  zero near an
interface. In order to calculate the spin polarization near the
boundary, Eq.(\ref{diffusion}) with $\mathbf{q}=-i\mathbf{\nabla}$
and $2\Gamma \delta(\mathbf{r})\delta_{ij}$ in the r.h.s. has to
be solved using appropriate boundary conditions. With the so
obtained $D^{ij}(\mathbf{r})$, the resultant spin density induced
by impurities placed at points $\mathbf{r}_i$ is given by
Eq.(\ref{sigma3})
\begin{equation}\label{sigmar}
S_j(\mathbf{r})=\sum_{n=x,y,z}\int d^2r^{\prime}
D^{jn}(\mathbf{r}-\mathbf{r}^{\prime})I^n_{tot}(\mathbf{r}^{\prime})\,,
\end{equation}
where the source term is obtained by the inverse Fourier transform
of (\ref{I2fin}):
\begin{eqnarray}\label{I3}
I^y_{tot}(\mathbf{r})&=& v_d N_F m^*\alpha
h_{k_F}^2\frac{1}{\Gamma^{2}n_i} \sum_i\delta(\mathbf{r}-\mathbf{r}_i) \nonumber\\
I^z_{tot}(\mathbf{r})&=&-v_d N_F h_{k_F}^2
\frac{1}{2\Gamma^{2}n_i}\sum_i\frac{\partial}{\partial
y}\delta(\mathbf{r}-\mathbf{r}_i)\nonumber\\
I^x_{tot}(\mathbf{r})&=&0\,,
\end{eqnarray}
where the relation $\Gamma^{\prime}=\Gamma/n_i$ is used because we
assumed that the target impurities are identical to the random
ones. The macroscopic polarization is obtained by averaging of
(\ref{sigmar}) and (\ref{I3}) over impurity positions. After
averaging over $x_i$ and the semiinfinite region $y_i>0$ the spin
polarization source (\ref{I3}) transforms to $I^n_{\text{av}}(y)$:
\begin{eqnarray}\label{I4}
I^y_{\text{av}}(y)&=& v_d N_F m^*\alpha
h_{k_F}^2\frac{1}{\Gamma^{2}} \nonumber\\
I^z_{\text{av}}(y)&=&-v_d N_F
h_{k_F}^2\delta(y-0^{+})\frac{1}{2\Gamma^{2}} \,.
\end{eqnarray}

It follows from (\ref{sigmar}) that the corresponding mean value
of the spin polarization $\mathbf{S}_{\text{av}}(y)$ satisfies the
diffusion equation (\ref{diffusion}) with the source $2\Gamma
I^n_{\text{av}}(y)$ in its rhs. The so obtained diffusion
equation, however, is not complete. One should take into account
that the boundary itself can create the interface spin
polarization. Most easily it can be done in the framework of the
Boltzmann approach. In terms of the Boltzmann function the spin
density is defined as
$\mathbf{S}_{\text{av}}(y)=\sum_{\mathbf{k}}\mathbf{g}_{\mathbf{k}}$,
and the charge density as $\sum_{\mathbf{k}}g_{\mathbf{k}}$. The
equation for the Boltzmann function can be written in the form
(see e.g. Ref. \onlinecite{Tang})
\begin{eqnarray}\label{qbg}
&&v_y \nabla_y \mathbf{g}_{\mathbf{k}} + 2(\mathbf{g}_{\mathbf{k}}
\times \mathbf{h}_{\mathbf{k}}) +
eE_x\frac{\partial \mathbf{g}^{(0)}_{\mathbf{k}}}{\partial k_x }\nonumber \\
&&= \frac{1}{\tau} \left(\mathbf{S}_E(y) -\mathbf{g}_{\mathbf{k}}
\right),
\end{eqnarray}
where $\mathbf{S}_E(y)=\delta(E-E_F)
\mathbf{S}_{\text{av}}(y)/N_F$ and
$\mathbf{g}^{(0)}_{\mathbf{k}}=-\mathbf{h}_{\mathbf{k}}\delta(E-E_F)$
is the equilibrium Boltzmann function. The terms proportional to
the charge component of the Boltzmann function have been omitted
in (\ref{qbg}) due to the system local electroneutrality, at least
in the scale of the mean free path, which is the smallest
characteristic scale of $\mathbf{g}_{\mathbf{k}}$ spatial
variations. The spin polarization source associated with the
boundary is given by a direct action of the electric field,
without taking into account secondary scattering from impurities.
Hence, the term with $\mathbf{S}_{\text{av}}(y)$ in r.h.s. of Eq.
(\ref{qbg}) can be ignored. Also, the boundary independent bulk
part of $\mathbf{g}_{\mathbf{k}}$ has to be subtracted from the
general solution of (\ref{qbg}). The so obtained interface
Boltzmann function will be denoted as
$\mathbf{g}_{\mathbf{k}\text{if}}$. The corresponding spin density
is
$\mathbf{S}_{\text{if}}(y)=\sum_{\mathbf{k}}\mathbf{g}_{\mathbf{k}\text{if}}$.
In order to calculate $\mathbf{g}_{\mathbf{k}\text{if}}$, Eq.
(\ref{qbg}) has to be supplemented with the boundary condition.
For a hard wall specularly reflecting boundary the condition is
simply
\begin{equation}\label{bc1}
\mathbf{g}_{k_x,k_y}|_{z=0}=\mathbf{g}_{k_x,-k_y}|_{z=0}\,.
\end{equation}
This condition means that the spin orientation does not change
after specular reflection from the interface. The solution of Eq.
(\ref{qbg}) satisfying (\ref{bc1}) can be easily found. As a
result, up to $o(\alpha^2)$ we obtain
\begin{eqnarray}\label{sif}
S^y_{\text{if}}(y)&=&S^x_{\text{if}}(y)=0 \nonumber \\
S^z_{\text{if}}(y)&=&8v_d\alpha^2\tau m^* \sum_{k_y
>0}k_y\delta(E_{\mathbf{k}}-E_F)e^{-y\frac{m^*}{k_y\tau}}\,.
\end{eqnarray}
Within the diffusion approximation the second of these equations represents a delta
source of the spin polarization with intensity
\begin{equation}\label{sourceif}
\frac{1}{\tau}\int_0^{\infty} dy S^z_{\text{if}}(y)=v_d N_F h_{k_F}^2 \frac{1}{\Gamma}\,.
\end{equation}
This source is exactly of the same magnitude, but opposite in sign to the
spin polarization emerging from impurities, that is represented by the integral of
$2\Gamma I^z_{\text{av}}(y)$, with $I^z_{\text{av}}(y)$ given by Eq. (\ref{I4}). Taking
into account that both sources are located at the interface, so that they cancel each
other, one sees that
only $y$-component of the source originating from impurity scattering retains
in the diffusion equation which acquires the form
\begin{eqnarray}\label{diffusion2}
\frac{\partial^2 S^z_{\text{av}}}{\partial y^2}-4m^{*}\alpha
\frac{\partial S^y_{\text{av}}}{\partial y} - 8m^{*2}\alpha^2 S^z_{\text{av}}&=&0 \nonumber \\
\frac{\partial^2 S^y_{\text{av}}}{\partial y^2}+4m^{*}\alpha
\frac{\partial S^z_{\text{av}}}{\partial y} - 4m^{*2}\alpha^2
S^y_{\text{av}} &=-&\frac{2\Gamma}{D} I^y_{\text{av}}\,.
\end{eqnarray}
The bulk solution of this equation is $S^z_{\text{av}}=0$ and
$S^y_{\text{av}}\equiv S_b=2\tau eE N_F \alpha$, that coincides
with the polarization obtained from
(\ref{sigmafin})-(\ref{sigmay}) at $q \rightarrow 0$.

In order to calculate the spin polarization near the interface we
employ the hard wall boundary conditions \cite{MalshStrip,
Accumulation4,Accumulation5} for (\ref{diffusion2}). Such boundary
conditions can be easily obtained from Eq. (\ref{qbg}) by
performing its summation over $\mathbf{k}$ and integrating from
$y=0$ to some point $y_0$, placed at the distance much larger than
$l$, but still small compare to $l_{so}$. A simple analysis of Eq.
(\ref{qbg}) shows that up to $o(\alpha^2)$ the sum over
$\mathbf{k}$ of the vector product in the l.h.s. of (\ref{qbg})
can be neglected, while the r.h.s and the term containing the
electric field turn to zero identically. As a result, we get
\begin{equation}\label{current}
\frac{1}{m^*}\sum_{\mathbf{k}}k_y \mathbf{g}_{k_x,k_y}|_{y=y_0}=
\frac{1}{m^*}\sum_{\mathbf{k}}k_y \mathbf{g}_{k_x,k_y}|_{y=0}\,.
\end{equation}
According to (\ref{bc}), the above sum is zero at $y=0$. Hence, it
is also zero at $y=y_0$. The latter sum coincides with the spin
current within its conventional definition \cite{Tang}, where a
contribution associated with the charge density due to the second
term of the velocity operator (\ref{v}) is ignored in an
electroneutral system. Using the gradient expansion of (\ref{qbg})
this current can easily be expressed \cite{Tang} through
$S^j_{\text{av}}|_{y=0}$, its $y$ derivative and the last term in
the l.h.s. of (\ref{qbg}). In this way one arrives to the boundary
conditions from Ref. \onlinecite{MalshStrip,
Accumulation4,Accumulation5}. We generalize these conditions by
adding possible effects of the surface spin relaxation (see also
Ref. \onlinecite{Accumulation1}). These additional terms are
characterized by the two phenomenological parameters $\rho_y$ and
$\rho_z$. Finally we obtain
\begin{eqnarray}\label{bc}
-D\frac{\partial S^z_{\text{av}}(y)}{\partial y}|_{y=0}&+&2D
m^{*}\alpha
(S^y_{\text{av}}(0)-S_b)=-\rho_z S^z_{\text{av}}(0) \nonumber \\
-D\frac{\partial S^y_{\text{av}}(y)}{\partial y}|_{y=0}&-&2D
m^{*}\alpha S^z_{\text{av}}(0)=-\rho_y S^y_{\text{av}}(0)\,.
\end{eqnarray}
One can easily see from (\ref{diffusion2}), (\ref{bc}) and
(\ref{bc}) that at $\rho_{x/y}=0$ the homogeneous bulk solution
$S^z_{\text{av}}=0, S^y_{\text{av}}(0)=S_b$ turns out to be the
solution of the diffusion equation everywhere at $y>0$. Therefore,
in this particular case the z-components of spin clouds from many
impurities completely cancel each other and there is no spin
accumulation near the interface, in agreement with Refs.
\onlinecite{MalshStrip,Accumulation2,Accumulation4,Accumulation5}.
At the same time, when $\rho_i\neq 0$ the out of plane component
of the spin density is not zero. In the case of weak surface
relaxation $\rho_i \ll D/l_{so}$ one obtains from
(\ref{bc})-(\ref{diffusion2})
\begin{equation}\label{S0}
S^z_{\text{av}}(0)=0.35 \rho_y \tau eE \frac{1}{2 \pi \hbar D} \,,
\end{equation}
where we inserted $\hbar$ to restore conventional units. It is
interesting to note that in such a regime of small enough $\rho_i$
the surface polarization does not depend on the spin-orbit
constant.

\section{spin-Hall resistance and energy dissipation}

As it was defined in the Introduction, the interface spin-Hall
resistance is given by
\begin{equation}\label{Rsh}
R_{sH}=\frac{S^z_{av}(0)}{N_F j} \,,
\end{equation}
where $j$ is the DC current density, $j=\sigma E$, with the Drude
conductivity $\sigma=ne^2\tau /m^{*}$. The so defined spin-Hall
resistance is closely related to the additional energy dissipation
which takes place due to spin accumulation and relaxation near
interfaces of a sample. Indeed, as was shown in Ref.
\onlinecite{MalshStrip}, the spin accumulation is associated with
a correction to the electric conductivity of DC current flowing in
the $x$ direction. For Rashba SOI the correction to the current
density has the form
\begin{equation}\label{deltaj}
\Delta j(y) = -\frac{e}{4m^{*}}\frac{\alpha^2
k_F^2}{\Gamma^2}\frac{\partial S^z_{av}}{\partial y}  \,.
\end{equation}
This expression is finite within the distance $\sim l_{so}$ from
the interface. After integration over $y$ one obtains a correction
to the electric current
\begin{equation}\label{deltaI}
\Delta I = \frac{e}{4m^{*}}\frac{\alpha^2
k_F^2}{\Gamma^2}S^z_{av}(0) \,.
\end{equation}
The corresponding interface energy dissipation (per the unit of
the interface length) can be expressed from (\ref{Rsh}) and
(\ref{deltaI}) as
\begin{equation}\label{w}
\Delta W=\Delta I E= \frac{m^{*}}{e\hbar}\alpha^2 \tau R_{sH} j^2
 \,.
\end{equation}
In its turn, $R_{sH}$ can be determined from Eq. (\ref{S0}). It
can be easily seen that $\Delta W >0$ if $\rho_y >0$.

\section{Results and Discussion}

Summarizing the above results, within the drift diffusion theory
we found out that the intrinsic spin-Hall effect induces in 2DEG a
nonequilibrium spin density around a spin independent isotropic
elastic scatterer. The z-component of this density has a shape of
a dipole directed perpendicular to the external electric field,
while the parallel to 2DEG polarization is isotropic. Due to the
D'yakonov-Perel' spin relaxation, the spin density decays
exponentially at a distance larger than the spin-orbit precession
length. Noteworthy, that such a cloud exists even in the case of
the Rashba spin-orbit interaction when the macroscopic spin
current is absent. We also calculated the macroscopic spin density
near an interface from taking the sum of clouds due to many
scatterers and independently averaging over their positions.
Surprisingly, in the case of the hard wall boundary, the so
calculated spin polarization exactly coincides with that found
from the drift diffusion or Boltzmann equations
\cite{MalshStrip,Accumulation2,Accumulation4,Accumulation5}. In
this case the out of plane component of the spin polarization is
zero, while the parallel polarization is a constant determined by
the electric spin orientation \cite{Edelstein}. Besides the hard
wall boundary we also considered a more general boundary condition
containing the interface spin relaxation, or the spin leaking
term. For such a general case $S^z \neq 0$. This polarization can
be associated with the local magnetic potential, because the
system attains its local equilibrium within the $S^z$ spatial
variation scale, which is much larger than $l$. The magnetic
potential, in its turn, is related to the DC electric current
density via the interface spin-Hall resistance. The latter was
shown to determine the additional energy dissipation due to
relaxation of the spin polarization near the interface.

Besides conventional semiconductor quantum wells, the results of
this work can be applied to metal adsorbate systems with strong
Rashba type spin splitting in the surface states
\cite{metalRashba}. In this case the spin cloud can be measured by
STM with a magnetic tip.

This work was supported by RFBR Grant No 060216699, NSC
95-2112-M-009-004, the MOE-ATU Grant and NCTS Taiwan. We are
grateful to the Centre for Advanced Study in Oslo for hospitality.

\end{document}